\documentclass[twocolumn]{aastex631}

\newcommand{\LIRA}{LIRA, Observatoire de Paris, Université PSL, Sorbonne Université, Université Paris Cité, CY Cergy Paris Université, CNRS, 92190 Meudon, France}

\shorttitle{Density structure radial evolution}
\shortauthors{Berriot et al.}

\graphicspath{{./}{figures/}}

\begin{document}

\title{Radial evolution of a density structure within a solar wind magnetic sector boundary}

\author{Etienne Berriot}
\affiliation{\LIRA}

\author{Pascal D\'emoulin}
\affiliation{\LIRA} 
\affiliation{Laboratoire Cogitamus, rue Descartes, 75005 Paris, France}

\author{Olga Alexandrova}
\affiliation{\LIRA}

\author{Arnaud Zaslavsky}
\affiliation{\LIRA}

\author{Milan Maksimovic}
\affiliation{\LIRA}

\author{Georgios Nicolaou}
\affiliation{Department of Space and Climate Physics, Mullard Space Science Laboratory, University College London, Dorking, Surrey RH5 6NT, UK}

\begin{abstract}

This study focuses on a radial alignment between Parker Solar Probe (PSP) and Solar Orbiter (SolO) on the 29$^{\text{th}}$ of April 2021 (during a solar minimum), when the two spacecraft were respectively located at $\sim 0.075$ and $\sim 0.9$~au from the Sun.
A previous study of this alignment allowed the identification of the same density enhancement (with a time scale of $\sim$1.5~h), and substructures ($\sim$20-30~min timescale), passing first by PSP, and then SolO after a $\sim 138$~h propagation time in the inner heliosphere.
We show here that this structure belongs to the large scale heliospheric magnetic sector boundary.
In this region, the density is dominated by radial gradients, whereas the magnetic field reversal is consistent with longitudinal gradients in the Carrington reference frame.
We estimate the density structure radial size to remain of the order L$_R \sim 10^6$~km, while its longitudinal and latitudinal sizes, are estimated to expand from L$_{\varphi, \theta} \sim 10^4$-$10^5$~km in the high solar corona, to L$_{\varphi, \theta} \sim 10^5$-$10^6$~km at PSP, and L$_{\varphi, \theta} \sim 10^6$-$10^7$~km at SolO.
This implies a strong evolution of the structure's aspect ratio during the propagation, due to the plasma's nearly spherical expansion.
The structure's shape is therefore inferred to evolve from elongated in the radial direction at $\sim$2-3 solar radii (high corona), to sizes of nearly the same order in all directions at PSP, and then becoming elongated in the directions transverse to the radial at SolO.
Measurements are not concordant with local reconnection of open solar wind field lines, so we propose that the structure has been generated through interchange reconnection near the tip of a coronal streamer.

\end{abstract}


\section{\label{sec:intro}Introduction}

The solar wind, resulting from the solar corona's expansion, carries with it the magnetic field of the Sun, creating sectors of different polarities within the Heliosphere.
During solar minima, the Sun's magnetic topology is nearly dipolar, with large coronal holes around the magnetic poles.
The solar wind distribution is therefore approximately bimodal, with slow  wind emerging from low latitudes regions and fast solar wind coming from the higher latitudes \citep[e.g.][]{McComas_2013_Ulysses_SW_latitudes_activity}.
At large scales, the interplanetary magnetic field \textbf{B} has two sectors of opposite polarities \citep{Wilcox_Ness_1965}.
The classical view of the boundary between those sectors is a large current sheet, the heliospheric current sheet (HCS), encased into a broader region of dense solar wind called the heliospheric plasma sheet \citep[HPS,][]{Winterhalter_1994_HPS}.

Studies of the solar wind near sector boundaries show 
that the HPS is a complex region formed of several substructures (sometimes identified as density blobs and magnetic flux ropes), possibly produced by magnetic reconnection at the tip of helmet streamers \citep{Crooker_1996_HPS_as_small-scale_transients, Crooker_2004_HPSs, Wang_1998_blobs_HCS-HPS, Sanchez-Diaz_2017_remote_blobs_HCS, Sanchez-Diaz_2019_in-situ_blobs_HCS, Liewer_2024_reconnection_HS_WISPR}.
These sectors boundaries are also increasingly dynamic when closer to the Sun, and often subject to magnetic reconnection \citep{Szabo_2020_HCS_PSP-WIND,Lavraud_2020_HCS,Phan_2021_Reconnection_HCS, Phan_2022_Reconnection_proton_beams_HCS}, also possibly contributing to the HPS complex structure.
The exact nature of the HPS (and HCS) is still debated, as are the processes occurring during its radial evolution within the heliosphere.
There are therefore different definitions of HCS/HPS employed in the literature.
Here, we refer to HCS as the thin region where the local magnetic field reverses, and true sector boundary (TSB) to where the global magnetic connectivity changes.
The HCS and TSB (associated with $\textbf{B}$ and  strahl electron pitch-angle reversals, respectively) are not necessarily collocated (see Section \ref{sec:density_structure_HPS-HCS} for more details).
We also use the term HPS to denote the broader region encasing the HCS and TSB, which separates the two magnetic sectors.

In this paper, we are interested in the radial evolution of the sector boundary by means of 2 points measurements.
Helios 1 \& 2 first enabled the possibility of studying the evolution of the same solar wind ``plasma parcel" by taking advantage of a configuration where two spacecraft are radially aligned \citep{Schwenn_line-ups_1981_a, Schwenn_line-ups_1981_b, Schwartz_Marsch_1983_radial_evolution}.
The recently launched Parker Solar Probe \citep[][PSP]{Fox_2016_Parker_Solar_Probe_science_goals} and Solar Orbiter \citep[][SolO]{Muller_2020_Solar_Orbiter_science_goals} are renewed opportunities to study the solar wind evolution during such radial alignment configurations \citep{Velli_2020_PSP_SolO, Telloni_2021}.
In order to infer when the same plasma is most likely to be crossed by both spacecraft, a new propagation method accounting for a constant solar wind acceleration was presented in \citet{Berriot_2024_propag}.
This allowed the identification of the same density structure passing through PSP ($\sim 0.075$~au) and SolO ($\sim 0.9$~au), for their radial alignment on the 29$^{\text{th}}$ of April 2021.

This structure identified in \citet{Berriot_2024_propag} was inferred to have radial gradients, and an estimation of its minimum latitudinal size was provided, however the broader physical context was not studied.
Here, we continue the study of \citet{Berriot_2024_propag}, with a focus on the more global physical context.
In Section \ref{sec:context}, we present the same density structure measured by PSP and SolO after a $\tau \sim 138$~h propagation time in the inner heliosphere, and give an estimation of its longitudinal extension.
In Section \ref{sec:density_structure_HPS-HCS}, we show that this density structure is closely related to the HCS and corresponds to a substructure of heliospheric plasma sheet.
Moreover, we infer that, while the density structure's gradients seem to be along the radial direction, the magnetic field reversal are more consistent with longitudinal gradients in the Carrington reference frame.
In Section \ref{sec:discussion}, we discuss the possible origins of the density structures and observed large scale plasma gradients.
Finally, we summarize the study and conclude in Section \ref{sec:conclusion}.


\begin{figure}[t!] 
\resizebox{\hsize}{!}{\includegraphics{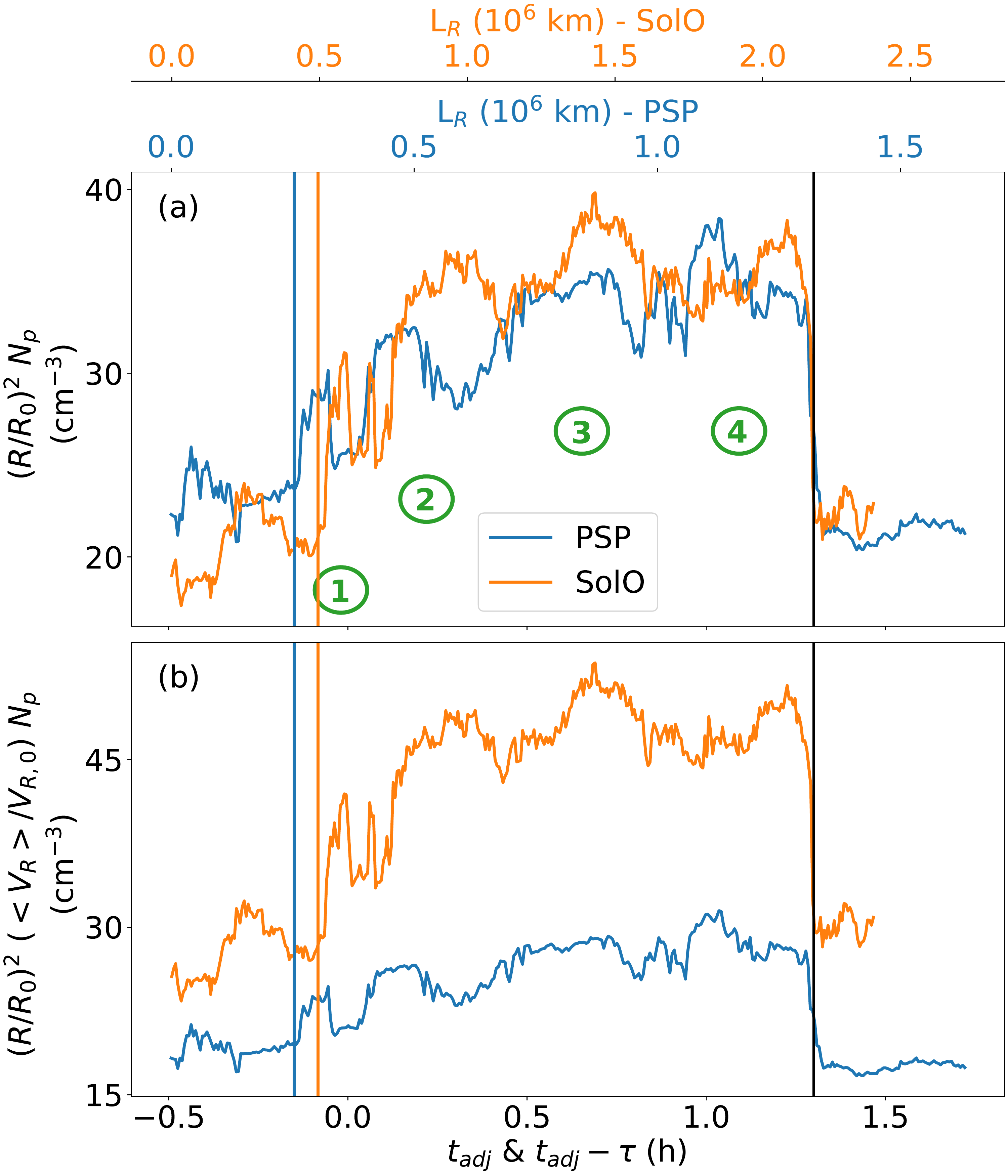}}
\caption{
\label{fig:structure_densite}
Proton density $N_p$ measured by PSP (blue curve) and SolO (orange curves) around the identified density structure, as functions of $t_{\rm adj}$  \& $t_{\rm adj} - \tau$ respectively, with $t_{\rm adj}$ defined in equation (\ref{eq:t_ajd}) and $\tau$ the plasma propagation time between PSP and SolO.
The top x-axis shows the inferred radial sizes of the structures ($L_{R}$) from measurements of both PSP (blue) and SolO (orange).
The principal substructures are labeled 1 to 4 on panel (a).
The measurements are averaged over 20~s to better highlight the substructures global shapes, and corrected to take into account the solar wind's spherical expansion (panels (a) and (b)) and acceleration (panel (b)).
Panel (b) is discussed in Section \ref{sec:plasma_compression}.
The start of the whole structure is indicated by blue and orange vertical lines for PSP and SolO, respectively, and the end of the structure is indicated by the same black vertical line for both PSP and SolO.
}
\end{figure}  

\section{\label{sec:context}Context and Density structure}

\subsection{In-situ measurements}

In this study, we use the velocity moments of ion distribution function measured by the instrument SPAN-Ion on PSP \citep{Livi_2021_SPAN-ion, Kasper_SWEAP_Instrument_PSP} and the PAS instrument from the SWA suite \citep{Owen_2020_SWA_SolO} on SolO.
The abundance of alpha particles relative to protons is estimated to be $\sim 1 \%$ during the studied time periods (not shown) for both spacecraft.
We therefore assume that the PAS moments we use here are representative of the proton population only.
This assumption is also relevant for PSP, but not needed as the data used here have already the alpha contribution filtered out.
Magnetic field data on PSP and SolO are from measurements of the FIELDS \citep{Bale_FIELDS_Instrument_PSP} and MAG \citep{Horbury_2020_MAG_SolO} instruments, respectively.
We also make use of the electron pitch-angle (PA) distribution measured by SPAN-E \citep{Whittlesey_2020_SPAN_electrons} on PSP, and EAS, also part of the SWA suite onboard SolO.

Trajectories in both, inertial, and corotating frames with the Sun are taken from NAIF SPICE kernels \citep{Acton_1996_NAIF} for the two spacecraft\footnote{These kernels are publicly available : \url{https://spdf.gsfc.nasa.gov/pub/data/psp/ephemeris/spice/} for PSP, and: \url{https://doi.org/10.5270/esa-kt1577e} for SolO.}.

\subsection{Same observed density structures at two solar distances}

We focus here on a radial alignment between PSP ($\sim 0.075$~au) and SolO ($\sim 0.9$~au) around the 29/04/2021.
We define a time origin $t_0$ (=~29/04/2021~00:45~UTC) when the two spacecraft are at the same longitude $\varphi$ in a heliocentric reference frame.
The time variable $t$ we consider is therefore
\begin{equation}
    t = t_{\rm UTC} - t_0
\end{equation}
where $t_{\rm UTC}$ is the Coordinated Universal Time.
Below, we summarize how we identified the same solar wind parcel in a previous study \citep{Berriot_2024_propag}.

In order to estimate the time interval corresponding to the plasma line-up, we began by modeling the solar wind propagation with a ballistic approximation.
We considered a constant solar wind acceleration constrained by the plasma bulk speed measurements on the two spacecraft, and effects of non-radial propagation (due to the formation of a stream interaction region (SIR), see Appendix \ref{sec:SIR}).
From this, we obtained that the most likely plasma line-up was for $t \sim 2$~h at PSP and for $t \sim 136$~h at SolO.
These were defined by determining the minimum distance between SolO and the plasma parcel's inferred position after its crossing by PSP, see \citet{Berriot_2024_propag} for further details.

\begin{table}
\caption{\label{tab:averages}Different physical quantities at PSP and SolO, averaged over the density structures intervals delimited by the vertical lines in Figure \ref{fig:structure_densite}.
The first three lines are plasma quantities (density $N_p$, magnetic field magnitude $B$ and bulk velocity $\mathbf{V}$).
The following lines provide spacecraft information (velocity $\mathbf{V}_{\rm SC}$, angular speed $\omega_{\rm SC}$, longitudinal angle relative to the plasma crossing $\varphi_{p, \rm SC}$, and distance to the Sun $R$).
Both $\mathbf{V}$ and $\mathbf{V}_{\rm SC}$ vectors components are given in the associated spacecraft local RTN frame.
}
\centering
\begin{tabular}{cll}
\hline \hline
Average Quantity & PSP & SolO \\
\hline
$Np$ (cm$^{-3}$) & 5530 & 42 \\
$B$ (nT) & 170 & 7 \\
$\mathbf{V}$ (km.s$^{-1}$) & (209, 3, -3) & (338, 12, 31) \\
$\mathbf{V}_{\rm SC}$ (km.s$^{-1}$) & (-23, 143, -7) & (5, 26, -1) \\
$\omega_{\rm SC}$ (rad.s$^{-1}$) & $1.25 \times 10^{-5}$ & $1.95 \times 10^{-7}$ \\
$\varphi_{p, \rm SC}$ (°) & -31 & -2 \\
$R$ (au) & 0.075 & 0.9 \\
\hline
\end{tabular}
\end{table}

To confirm and precise the above estimation, we searched for a same structure, used as a marker for the identification of what can be considered the same plasma parcel passing through both spacecraft.
A visual inspection, coupled with a cross-correlation method, indeed allowed us to find what we believe to be the same density enhancement on PSP and SolO for a propagation time of $\tau = 137.6$~h.
This structure is shown in Figure \ref{fig:structure_densite}(a).
Table \ref{tab:averages} summarizes several solar wind plasma parameters on both spacecraft, averaged over time intervals corresponding to the density structures and delimited by vertical lines in Figure \ref{fig:structure_densite}.

In Figure \ref{fig:structure_densite}, the measured proton density $N_p$ is plotted as a function of the adjusted time $t_{\rm adj}$ (defined below with Equations (\ref{eq:dt_ajd}) and (\ref{eq:t_ajd})) for PSP (blue) and as a function of $t_{\rm adj} - \tau$ for SolO (orange).
To account for a spherical expansion of the plasma, the density has been corrected on both panels by a factor $(R/R_0)^2$, with $R_0 =1$~au.

Comparing the measurements as functions of time $t$ is the most pertinent approach if the different substructures have similar outward accelerations, as shown by \citet{Berriot_2024_propag}.
However, because the spacecraft have a non-zero radial speed during the studied interval, the time variable $t$ should be adapted to take this into account.
In order to do so, we define $dt_{\rm adj}$ an adjusted time element between two subsequent measurements (at times $t$ and $t+dt$) as:
\begin{equation}
    dt_{\rm adj}(t) = \frac{V_R(t) - V_{R, \rm SC}(t)}{V_R(t)} dt, \label{eq:dt_ajd}
\end{equation}
with $V_R (t)$ and $V_{R, \rm SC}(t)$ the proton and  spacecraft radial speed, respectively.
Considering purely advected plasma structures, $(V_R - V_{R, \rm SC})dt$ is the radial extension of the solar wind explored during $dt$, and $dt_{\rm adj}$ is the corresponding time interval if the spacecraft had no radial speed.
This allows to define a new adjusted time variable $t_{\rm adj}$ as:
\begin{eqnarray}
    t_{\rm adj}(t) = t_{0, \rm adj} + \int_{0}^t \frac{V_R(t') - V_{R, \rm SC}(t')}{V_R(t')} dt', \label{eq:t_ajd}
\end{eqnarray}
where $t'$ is the integration variable.
We set $t_{0, \rm adj}=0$~h for PSP, while for SolO $t_{0, \rm adj}$ is set such that the second boundary of the structure (vertical black line in Figure \ref{fig:structure_densite}) matches between the two spacecraft, as to remain consistent with \citet{Berriot_2024_propag}.

\begin{figure*}[ht!] 
\resizebox{\hsize}{!}{\includegraphics{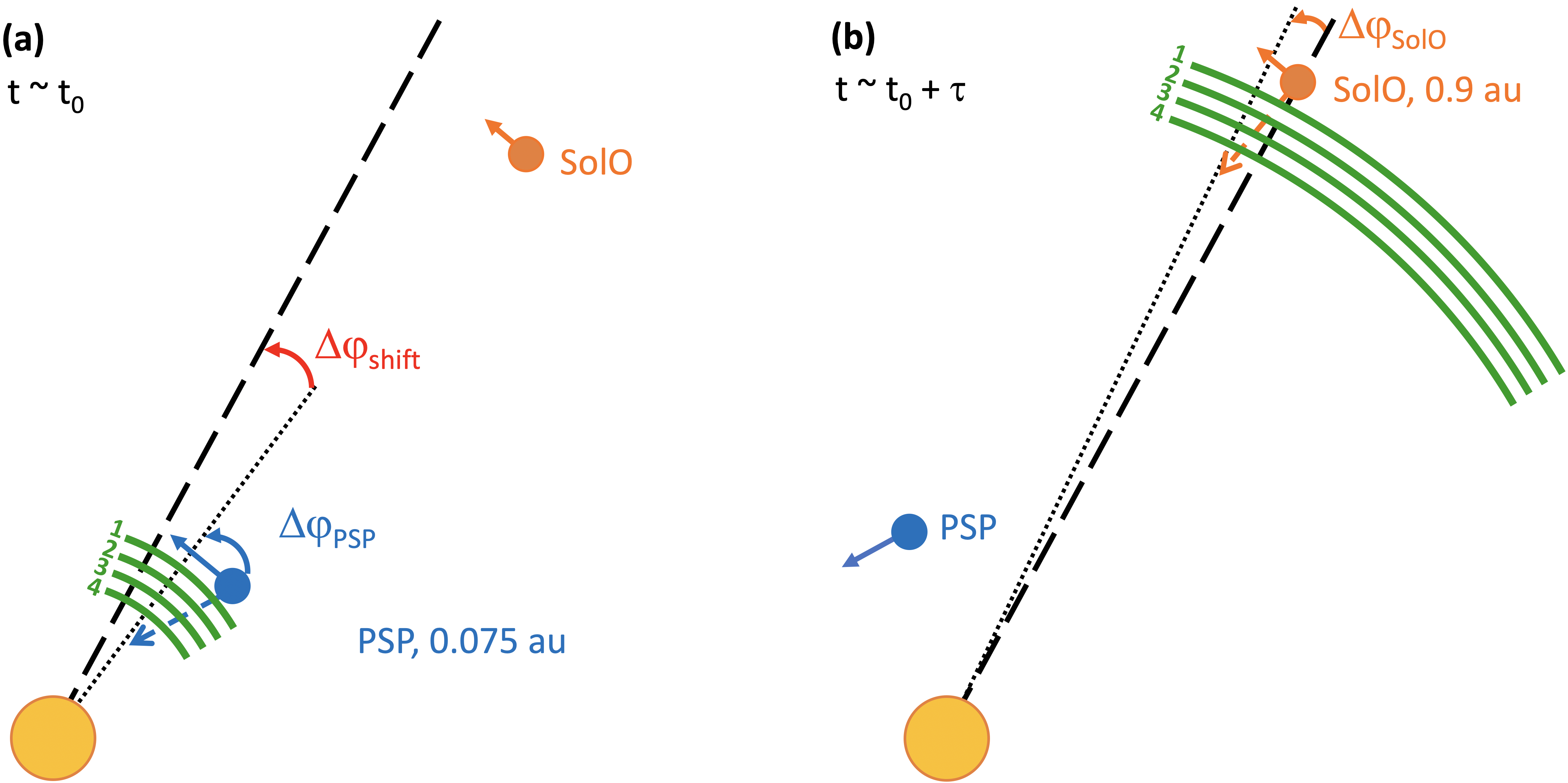}}
\caption{\label{fig:schematic_structures_inertial}
Not to scale schematic of the density structure's minimal extension inferred from PSP and SolO measurements, in an inertial reference frame.
For clarity reasons, the structure's longitudinal extension is exaggerated by approximately an order of magnitude.
The density structure with radial gradients is depicted by four green arcs labeled 1 to 4 as in Figure \ref{fig:structure_densite}, before its passage at PSP in (a) and before its passage passage at SolO in (b).
PSP and SolO positions are indicated in blue and orange, respectively, and their velocity vector by plain arrows of matching colors.
PSP and SolO trajectories within the structure are shown with blue and orange dashed arrows, respectively.
The thick black dashed line marks the radial between the Sun and SolO position at the leading time observation of the structure by SolO.
The thin black dotted line is the radial between the Sun, and the spacecraft (PSP in (a), SolO in (b)) just after its passage of the density structure.
The longitudes $\Delta \varphi_{\rm shift}$ is the angular extension between PSP exit and SolO entrance in the structure.
The scanned longitudes  $\Delta \varphi_{\rm PSP}$ and  $\Delta \varphi_{\rm SolO}$ within the density structure are in blue and orange, respectively.
The minimal angular extension of the structure $\Delta \varphi_{\rm str}$ is indicated in purple on panel (b).
}
\end{figure*}  

This correction  is needed to compare precisely the measurements of the two spacecraft (which have different radial motions).
The plasma radial speed in the local spacecraft rest frame is $V_{R} - V_{R, \rm SC}$, so a $V_{R, \rm SC} < 0$ implies a shorter time of passage, and $V_{R, \rm SC} > 0$ a larger time of passage, than if the spacecraft had no radial speed.
Here, $V_{R, \rm SolO} \approx 5$~km/s, and $V_{R, \rm PSP} \approx - 23$~km/s, so that mainly PSP alters $t_{\rm adj}$ as compared to $t$.
Still, because $V_{R, \rm PSP} / V_{R} \simeq 1/10$, this effect do not challenges the structure and substructures association, as seen by comparing Figure \ref{fig:structure_densite}(a) to Figure 8(a) in \citet{Berriot_2024_propag}, where measurements are plotted as functions of $t$.

Estimates of the structures radial sizes L$_{R}$ at PSP (blue) and SolO (orange) are shown on the top-x axis of Figure \ref{fig:structure_densite}.
Those were calculated using $t_{\rm adj}$ and the average recorded proton bulk speed $V_R$ at both spacecraft, with the origin at the beginning of the shown time intervals (L$_{R}(t_{\rm adj}=-0.5$ h$) = 0$).
Although they are of the same order of magnitude, the L$_{R}$ estimated at SolO ($\sim 1.5 \times 10^6$~km) is larger than the one estimated at PSP ($\sim 10^6$~km) due to the solar wind acceleration.
Thus, the structures are slightly stretched radially from PSP to SolO \citep{Berriot_2024_propag}.

We see on Figure \ref{fig:structure_densite}(a) that the structure has been very well conserved despite its $\sim 0.8$~au journey in the inner heliosphere.
We are even able to associate the principal substructures (labeled 1 to 4) within it.

PSP and SolO are scanning the solar wind with very different orbital speeds.
In the solar wind local reference frame, the spacecraft velocity is $\textbf{V}'_{\rm SC} = \textbf{V}_{\rm SC} - \textbf{V}$, with $\textbf{V}$ and $\textbf{V}_{\rm SC}$ the plasma and spacecraft velocity, respectively.
Each spacecraft therefore scans the plasma with a longitudinal angle $\varphi_{p, \rm SC} = \arctan(V'_{T, \rm SC}/V'_{R, \rm SC})$, where we considered that, due to PSP and SolO trajectories being near the same plane ($\sim 3$° difference), $V'_{T, \rm SC} \simeq V'_{\varphi, \rm SC}$.
The average of this angle over the identified density structure (given in Table \ref{tab:averages}) is $\langle \varphi_{p, \rm PSP} \rangle \simeq -31$° and $\langle \varphi_{p, \rm SolO} \rangle \simeq -2$°, for PSP and SolO, respectively.

The fact that there is such a temporal correspondence of $N_p(t)$ in Figure \ref{fig:structure_densite} (a), despite the very different $\varphi_{p, \rm SC}$, is only possible if the density enhancement has dominant radial gradients.
Otherwise, variations due to non-radial gradient would be visible on PSP  measurements but mostly not on SolO, because of the different spacecraft velocities with respect to the plasma flow.
These gradients are measured as the solar wind carries the density structure, with a mostly radial velocity.
The good matching between the density measurements taken by PSP and SolO therefore not only indicates that the structure had undergone weak evolution during its travel, but it also implies that the density radial gradients are dominant along the passage of the two spacecraft.

\begin{figure*}
\includegraphics[width=1\textwidth]{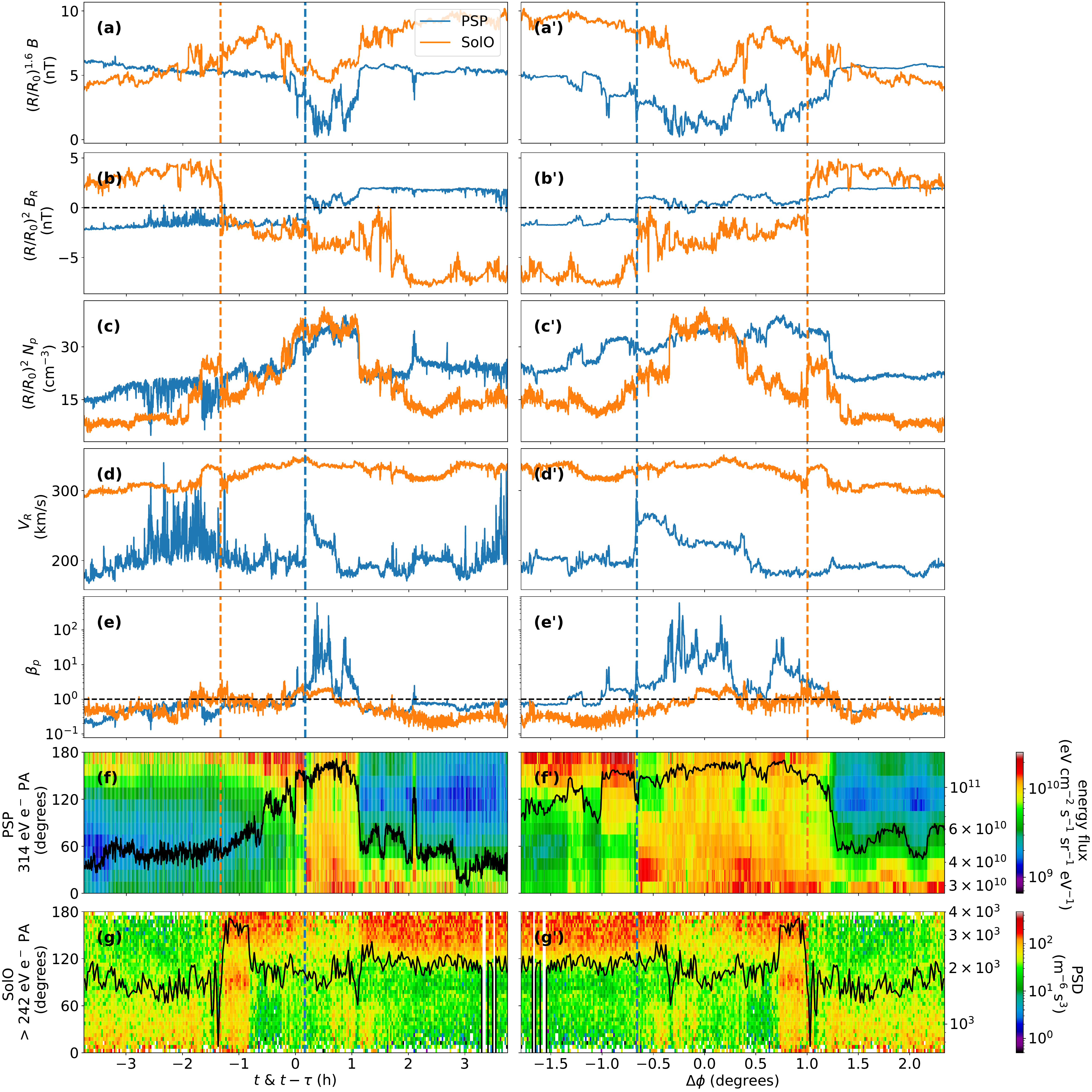}
\caption{\label{fig:HCS-HPS_PSP-SolO} Measurements taken by PSP (blue) and SolO (orange) around the structures shown in Figure \ref{fig:structure_densite}, with the position of the HCS (and coincident TSB) marked by matching colors vertical dashed lines.
On the left panels, data are plotted as functions of $t$ for PSP and $t - \tau$ for SolO, while on the right panels they are  plotted as functions of the relative longitudinal variation $\Delta \phi$.
The time origin is set at $t_0$=~29/04/2021~00:45~UTC and the propagation time is $\tau = 137.6$~h.
Panels (a) \& (a') show the magnetic field's magnitude $B$ and panels (b) \& (b') show its radial radial component $B_R$.
The proton density $N_p$ is on panels (c) \& (c'), the proton radial velocity $V_R$ on panels (d) and (d'), and the proton beta on panels (e) and (e').
To take into account the plasma's nearly spherical expansion, $B_R$ and $N_p$ are corrected by $(R/R_0)^2$, and $B$ by $(R/R_0)^{1.6}$.
The energy flux of strahl electrons is shown in colors (scale on the right side) as a function of time and Pitch-Angle (PA) on panels (f) \& (f') for PSP and (g) \& (g') for SolO where PSD stands for phase space density.
The energy flux value summed over all the angles is shown by a black line on the same panels, with the scale indicated on the right axis.
}
\end{figure*}

\subsection{\label{sec:plasma_compression}Plasma compression with distance}

On Figure \ref{fig:structure_densite}(b), in addition to the $(R/R_0)^2$ correction, measurements are also corrected by $\langle V_R \rangle / V_{R,0}$ to account for the plasma's acceleration, with $\langle V_R \rangle$ the radial proton speed measured by each spacecraft, averaged on the shown intervals (see Table \ref{tab:averages}), and an arbitrary speed reference $V_{R,0} = 250$ km/s.
For a steady state, the solar wind radial acceleration should induce a proportional decrease of $N_p$, as given by the continuity equation for a purely spherical expansion:
\begin{equation}
    \partial_R(N_p V_R R^2) = 0 \Rightarrow N_p \propto 1/(R^{2} V_R). \nonumber
\end{equation}
Surprisingly, the corrected $N_p$ at SolO is $\sim 1.5$ times higher than what expected from a steady state radial propagation.
This implies a compression of the structure by the same factor.
This compression is most likely due to the formation of an SIR during the plasma propagation (Appendix \ref{sec:SIR}).
Some contribution might also be linked to a different calibration of the spacecraft instruments.
Indeed, the proton densities can also be compared with electron density values in the density structure (not shown here), estimated by quasi-thermal noise (QTN) spectroscopy \citep[see][and references therein]{Meyer-Vernet_2017_QTN_review}.
For PSP, the electron densities are very close to SPI proton's density value \citep[see][for QTN using FIELDS data]{Moncuquet_2020_PSP_QTN}, while for SolO, estimations from RPW data \citep{Maksimovic_2020_RPW, Khotyaintsev_2021_density_RPW} give a factor $\sim 1.1$-$1.2$ lower than PAS.
The compression factor of 1.5 might therefore be a slight overestimation of the real compression effect from the SIR, which would be closer to $1.3$ according to QTN data.


\subsection{\label{sec:str_inertial_long_ext}Structure longitudinal extension}

In the following, we assume a spherical expansion of the solar wind, and a non decreasing of the angular size of the density structures during this expansion.
The crossing of the same density structure by the two spacecraft allows the estimation of its minimum longitudinal extension  $\Delta \varphi_{\rm str}$, in an inertial reference frame, taken here as the Heliocentric Inertial (HCI) reference frame.
This extension comes from the contribution of 3 terms and can be written as:
\begin{equation}\label{eq:delta_phi_contrib}
     \Delta \varphi_{\rm str} =  \Delta \varphi_{\rm PSP} + \Delta \varphi_{\rm SolO} + \Delta \varphi_{\rm shift},
\end{equation}
where $\Delta \varphi_{\rm PSP}$ and $\Delta \varphi_{\rm SolO}$ are the longitudes covered by PSP and SolO during their crossing of the structure, respectively.
The contribution $\Delta \varphi_{\rm shift}$ is the longitude shift between the two spacecraft observations of the structure. 
A schematic is depicted on Figure \ref{fig:schematic_structures_inertial}.

Both spacecraft have $\Delta \varphi_{\rm SC} > 0$ as they orbit in the +T direction, and $\Delta \varphi_{\rm shift} > 0$ since SolO cross the structure at a higher longitude than PSP.
This implies that Equation (\ref{eq:delta_phi_contrib}) can be simplified as:
\begin{equation}
    \Delta \varphi_{\rm str} = \varphi_{\rm SolO}(t=t_{\rm end, SolO}) - \varphi_{\rm PSP}(t=t_{\rm start, PSP}),
\end{equation}
where $\varphi_{\rm PSP}(t=t_{\rm start, PSP})$ is the longitude of PSP when it enters the structure, and $\varphi_{\rm SolO}(t=t_{\rm end, SolO})$ the longitude of SolO when it exits the structure.
For the studied case ($t_{\rm start, PSP}$~=~29/04/2021~00:34~UTC and  $t_{\rm end, SolO}$~=~04/05/2021~19:29~UTC), we obtain $ \Delta \varphi_{\rm str} \simeq 6.5$°, see Appendix \ref{sec:appendix_phi_r} for more details.

Deflection of the plasma (due to a SIR here) in the longitudinal direction reduces this estimation by an amount $\Delta \varphi_{\rm NR} \simeq 2.0$° (the subscript "NR" stands for "non-radial", see Appendix \ref{sec:appendix_phi_nr}).
This corresponds to a spatial extension $\Delta \varphi_{\rm str}^* = \Delta \varphi_{\rm str} - \Delta \varphi_{\rm NR} \simeq 4.5$°.
So, the lower bound longitudinal length (estimated as arcs of angular extension $\Delta \varphi_{\rm str}^*$) of the density structures is about $L_{\rm \varphi, PSP} \simeq 9 \times 10^5$~km at PSP, and about $L_{\rm \varphi, SolO} \simeq 1.1 \times 10^7$~km at SolO.

We note that the above calculations are valid in this configuration, because the structure is crossed by PSP before the predicted plasma line-up (see \citet{Berriot_2024_propag}).
In the case where the same structure is crossed by the inner spacecraft after the predicted plasma line-up, the minimal extension would be
\begin{equation}
    \Delta \varphi_{\rm str} = \varphi_{\rm PSP}(t=t_{\rm end, PSP}) - \varphi_{\rm SolO}(t=t_{\rm start, SolO}).
\end{equation}
In the case of a crossing around the predicted plasma line-up, the outer spacecraft would only be sampling longitude regions already crossed by the inner spacecraft (because $\omega_{\rm PSP} \gg \omega_{\rm SolO}$).
Therefore, estimation of the structure's minimal longitudinal extension would reduce to
\begin{equation}
    \Delta \varphi_{\rm str} = \varphi_{\rm PSP}(t=t_{\rm end, PSP}) - \varphi_{\rm PSP}(t=t_{\rm start, PSP}).
\end{equation}

\section{\label{sec:density_structure_HPS-HCS}Density structure and sector boundaries}

\subsection{In-situ observations as functions of time}

In order to study the solar wind context, we now consider measurements of magnetic field, and suprathermal electron pitch-angle (PA) distribution for a longer time interval around the identified density structure.
On the left panels of Figure \ref{fig:HCS-HPS_PSP-SolO} (a-g), we show the physical parameters measured by PSP and SolO in and around the structures, for $t \; \& \; t-\tau \in [-3.75, 3.75]$~h.
The density structures time intervals also correspond to a local decrease in the magnetic field's magnitude $B$ (a), as already pointed in \citet{Berriot_2024_propag}.

From Figure \ref{fig:HCS-HPS_PSP-SolO}, we can see that the density structure in Figure \ref{fig:structure_densite} corresponds to a substructure of the heliospheric plasma sheet (HPS).
At PSP, the density structure covers the whole observed HPS, while at SolO, the magnetic field inversion corresponding to the HPS ($t-\tau \in [-1.3, 1.7]$~h) lasts longer than than the density structure observation.
Here, we define the HPS as the broad buffer region between the two large scale heliospheric magnetic sectors, and encompassing the thinner heliospheric current sheet (HCS).
The HPS typically has a denser plasma, a lower magnetic field, leading to a higher plasma beta than that of its surroundings \citep[e.g.][]{Winterhalter_1994_HPS}, although the precise nature of this region is still not yet fully understood.

Simultaneous abrupt reversals of the magnetic field's radial component $B_R$, Figure \ref{fig:HCS-HPS_PSP-SolO}(b), and of the strahl electron ($\sim 314$~eV) PA energy flux are present for both PSP (f) and SolO (g).
These reversals of $B_R$ and strahl electron PA, indicating crossings of the Heliospheric Current Sheet (HCS) and True Sector Boundary (TSB), respectively, are marked by vertical dashed lines of matching colors for the two spacecraft.
Here, we are therefore in the simple case where measurements of the local magnetic field are in agreement with the global magnetic connectivity to the Sun inferred from the strahl electrons PA.
Previous observations have shown that this is not always verified as the TSB and HCS are not necessarily coincident depending on the sector boundaries magnetic field's topology \citep{Kahler_1994_SB_PA, Crooker_2004_HCS-HPS_TSB, Foullon_2009_multi-layered_HCS}.

The magnetic field and PA reversals indicate that the sector boundaries were crossed in opposite directions by the two spacecraft. Indeed, PSP went from a sector with inward directed magnetic fields lines ($B_R < 0$~\&~PA~$\sim 180$°) to a sector with outward directed magnetic fields lines ($B_R > 0$~\&~PA~$\sim 0$°), and the opposite for SolO.
The expected evolution of the solar wind during its propagation could not explain this behaviour.
In fact, due to its proximity to the Sun for the studied time intervals, PSP's orbit around the Sun (in an inertial reference frame) is faster than the Sun's rotation.
This is not the case for SolO which is further away from the Sun, implying that the two spacecraft are scanning solar wind sources in opposite directions.

\begin{figure*}[ht!] 
\resizebox{\hsize}{!}{\includegraphics{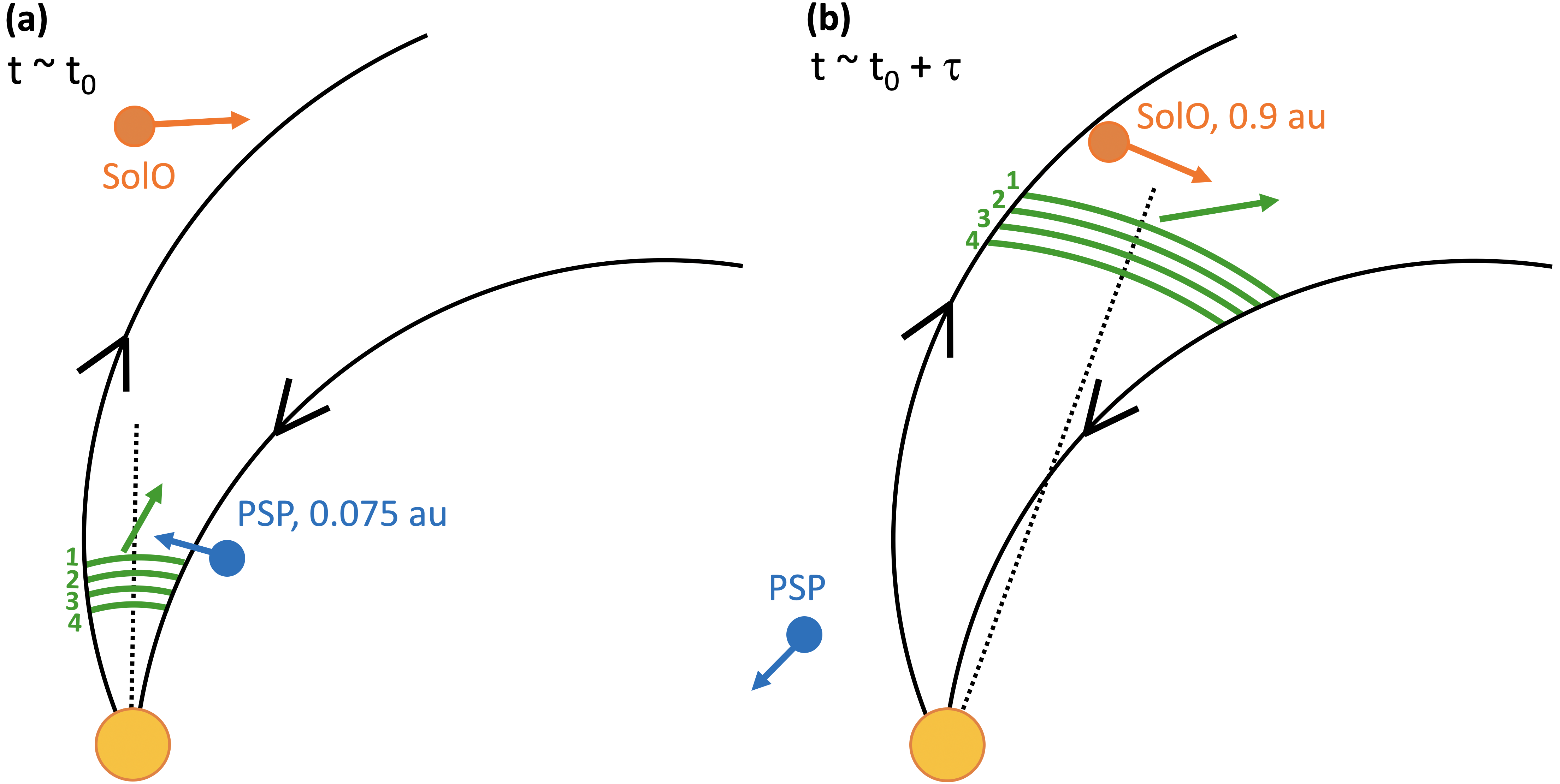}}
\caption{\label{fig:schema_structures} Schematic interpretation of the HPS crossing by PSP and SolO for the studied alignment, represented as an ecliptic cut in the Carrington reference frame.
The density structure with radial gradients is depicted, at two different times, by four green arcs labeled 1 to 4 as in Figure \ref{fig:structure_densite}.
This structure represents a part of the HPS.
The inward and outward magnetic field lines bordering the density are drawn as black plain lines with arrows.}
\end{figure*}  

\subsection{\label{sec:Carrington_longitude} In-situ observations as functions of longitude}

In order to take into account the relative angular evolution of the spacecraft with respect to the Sun, we now consider PSP and SolO measurements as functions of $\Delta \phi$, the spacecraft longitude variation in the Carrington reference frame.
Considering an angular rotation speed around the Sun $\omega_{\rm SC}$ for a spacecraft SC (PSP or SolO here), a time interval $\Delta t$ can be converted in the relative longitudinal variation $\Delta \phi$ as:
\begin{equation}\label{eq:delta_phi_carr_omega}
    \Delta \phi = (\omega_{\rm SC} - \omega_{\rm Sun}) \Delta t
\end{equation}
where $\omega_{\rm Sun}$ is the solar rotation angular frequency.
Since $\omega_{\rm PSP} > \omega_{\rm Sun}$, this $\Delta \phi$ increases over time for PSP, whereas $\Delta \phi$ decreases over time for SolO because $\omega_{\rm SolO} < \omega_{\rm Sun}$.

In the following, $\Delta \phi$ is evaluated directly from the spacecraft trajectories in the Carrington reference frame rather than using Equation (\ref{eq:delta_phi_carr_omega}).
However, by comparing data using $\Delta \phi$, we loose the direct association, through $\tau$, between the measurements at the two spacecraft. We therefore use the density structures shown in Figure \ref{fig:structure_densite} to set a common longitude origin.
More precisely, we define:
\begin{equation}
    \Delta \phi (t) = \phi_{\rm SC}(t) - \phi_{0, \rm SC} \label{eq:delta_phi_carr_SC}
\end{equation}
where $\phi_{\rm SC}(t)$ is either the longitude of PSP ($\phi_{\rm PSP}(t)$) or SolO ($\phi_{\rm SolO}(t)$) as seen in the Carrington reference frame, and $\phi_{0, \rm PSP} = \phi_{\rm PSP}(t=0.5$~h) \& $\phi_{0, \rm SolO} = \phi_{\rm SolO}(t-\tau=0.5$~h) as these times roughly correspond to the center of the density structures.

On the right panels of Figure \ref{fig:HCS-HPS_PSP-SolO}(a'-g') are the PSP and SolO measurements as functions of $\Delta \phi$ for the same parameters as those shown in the left panels (a-g).
SolO measurements have been taken for the same time interval than previously ($t-\tau \in [-3.75, 3.75]$~h), corresponding to $\Delta \phi \in \sim [-1.8, 2.3]$ degrees.
This range of $\Delta \phi$ defines the associated limits of the time interval at PSP ($t \in [-0.4, 1.7]$~h).
As expected, the density enhancement measurements do not match anymore between the two spacecraft (Figure \ref{fig:HCS-HPS_PSP-SolO}(c')).
This is consistent with this structure having radial gradients.

 \subsection{\label{sec:mag_inversion}Magnetic field inversion in the HPS}

In the right panels of Figure \ref{fig:HCS-HPS_PSP-SolO}, the magnetic field's radial component $B_R$ (b') and electron PA (f', g') reversals are now done in the same direction.
We remark that the HPS magnetic field inversion looks like a bifurcated current sheet on the two spacecraft, often associated with magnetic reconnection events close to the Sun \citep{Phan_2022_Reconnection_proton_beams_HCS, Eriksson_2022_HCS_avalanche_reconnection}.
At PSP, there is a sharp reversal of $B_R$ and PA (at $\Delta \phi \approx - 0.7$°, blue vertical dashed line), followed by a region with a small mostly positive $B_R$ and small magnetic field magnitude $B$ (from 300-350~nT outside the HPS, to 25-50~nT in regions where it is the lowest).
At SolO, the HPS has a negative magnetic polarity $B_R$ and is bordered by two intense current sheets, with one of them also corresponding to the HCS (located at $\Delta \phi \sim 1$°, orange vertical dashed line) and coincident TSB.

The HPS magnetic field reversal extends over $\Delta \phi \in [-0.7, 1.3]$° for PSP and $\Delta \phi \in [-0.8, 1.0]$° for SolO.
Therefore, the HPS has kept a comparable longitudinal extension ($\sim 2.0$° at PSP and $\sim 1.8$° at SolO) within the heliosphere during the solar wind's expansion.
The fact that this extension is slightly shorter at SolO can be explained by the SIR formation, see Appendix \ref{sec:SIR}.

The solar wind strahl electrons are considered to be produced at the Sun and to then be traveling freely along the heliospheric magnetic field lines.
Measurements by the two spacecraft therefore indicate that the magnetic field lines remain linked to the Sun throughout the HPS crossings as no large scale strahl electron flux dropout is observed (see solid black line in Figure \ref{fig:HCS-HPS_PSP-SolO} (f',g')).

The whole HPS crossing by PSP also coincides with a more isotropic and diffuse PA distribution, coincident with the decrease in the magnetic field's magnitude $B$.
In this physical context, the strong $B$ decrease implies a weaker relevance of the privileged direction induced by the magnetic field, hence leading to a more isotropic plasma.
Mechanisms contributing to the PA isotropization possibly include scattering of the strahl electrons by instabilities, triggered more easily in the higher $\beta$ regions \citep[e.g.][and references therein]{Halekas_2021_near-Sun_electron_heat_flux}.
Counter streaming electrons are also observed by PSP within $\Delta \phi$ = [0.5°,1.1°].
They are possible markers of field lines connected to the Sun at both ends.
At SolO, the measured $\beta$ is lower, and the strahl electron PA also shows a very diffuse region, but more localized on the side of the HCS ($\Delta \phi \in [0.8, 1]$°).
The rest of the sector boundary crossing exhibits some deviations from the unidirectional electron PA distribution (especially for $\Delta \phi \in [-0.5, 0]$°).
This is however linked to the localized magnetic field's structuring and goes beyond the scope of this study.


\begin{figure*}[ht!] 
\resizebox{\hsize}{!}{\includegraphics{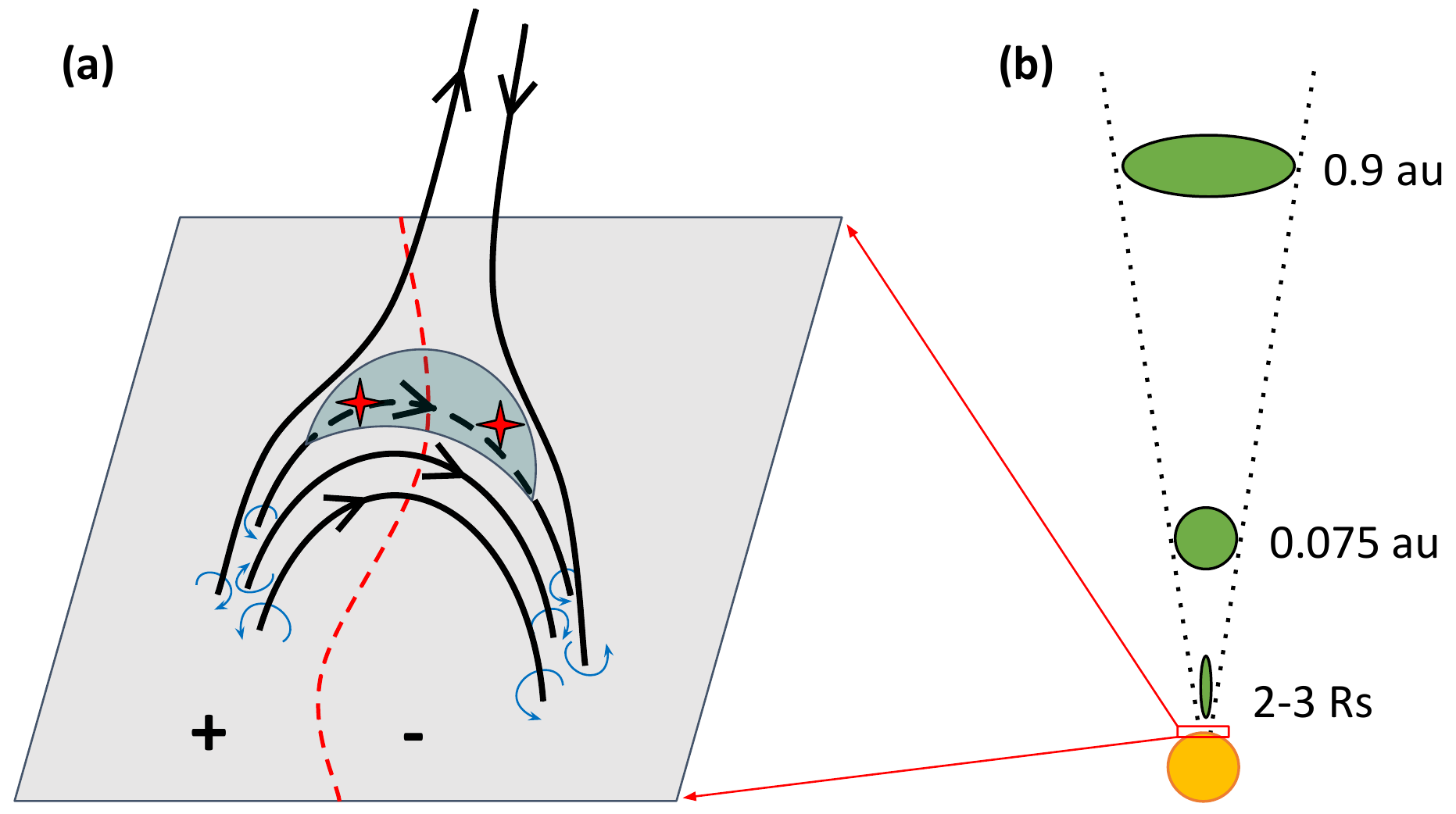}}
\caption{\label{fig:schema_interchange} Panel (a) shows a simplified representation of a coronal streamer with open and closed field lines (black) above a bipolar region (inversion line shown with dashed red line).
The current sheet, between open and closed field lines, before the onset of interchange reconnection is depicted as the transparent blue shape.
Expected reconnection regions are indicated by the two red stars.
Photospheric motion are represented as curved blue arrows around the field line footpoints.
Panel (b) is a schematic (not to scale) of the density structure overall shape (green) in the heliosphere at three different times and distances from the Sun.
}
\end{figure*}  

\subsection{Interpretation}

Figure \ref{fig:schema_structures} shows a schematic and simplified interpretation of the studied case, where we considered a cut in the ecliptic plane as viewed within the Carrington reference frame.
A density structure (and associated substructures) with radial gradients, depicted by 4 green arcs, passes through PSP around a time $t_0$ and through SolO around $t_0 + \tau$.
This structure is part of the HPS.

Due to the spacecraft trajectory in this reference frame, the HPS (and HCS) are crossed in opposite direction by PSP and SolO, when comparing measurements as functions of time (Figure \ref{fig:HCS-HPS_PSP-SolO}, left panels).
Therefore, measurements of magnetic field are better matching when compared as functions of longitude because the magnetic field inversion is linked to the source region on the Sun.
On the other hand, the density structure has dominant radial gradients, and is advected with the solar wind, which explains why there is a better correspondence when comparing $N_p$ measurements as functions of time.

We also remark in Figure \ref{fig:HCS-HPS_PSP-SolO} that the TSB and HCS are present within the density structure at PSP (blue vertical dashed line), and outside of it at SolO (orange vertical dashed line).
The HPS is seen with a mostly positive polarity ($B_R > 0$~\&~PA~$\sim 0$°) at PSP, with the HCS lying on the HPS side with $\Delta \phi < 0$, while at SolO, the HPS has a mostly negative polarity ($B_R < 0$~\&~PA~$\sim 180$°), with the HCS lying on the other HPS side ($\Delta \phi > 0$).
Although it already has been observed that the HCS tends to be located at one edge of the HPS \citep[e.g.][]{Winterhalter_1994_HPS}, it is trickier to explain the above HCS displacement and of polarity.

A possible explanation is that this effect is due to some local inclination of the HCS.
There is a latitude difference of $\Delta \theta \sim 3$° between PSP and SolO around their radial coalignment.
Although the deflection due to the SIR is estimated to bring the plasma observed by PSP closer to SolO's latitude \citep{Berriot_2024_propag}, it isn't enough to cover the entirety of $\Delta \theta$, and there is also a shift in longitude between the spacecraft and the predicted plasma line-up (Section \ref{sec:str_inertial_long_ext} and Appendix \ref{sec:appendix_phi}).
PSP and SolO are thus crossing the density structures and HCS at two different locations.
Furthermore, the two spacecraft also orbit at different speeds around the Sun and therefore cross the density structures with different trajectories (see Section \ref{sec:str_inertial_long_ext} and Figure \ref{fig:schematic_structures_inertial}).
Then, the spacecraft cross the density structure at two
significantly different locations.


\section{\label{sec:discussion}On the origin of the density structures}

\subsection{Generation by reconnection in the solar wind}

We are now interested in the generation mechanism of this density structure and substructures.
A first explanation would be that the structure has been produced by reconnection of the HCS open magnetic field lines.
We see on Figure \ref{fig:HCS-HPS_PSP-SolO} (d) and (d') that an important $+V_R$ jet is present in PSP data ($t\in [0.15, 0.75]$~h and $\Delta \phi \in [-0.7, 0.5]$°).
This could be a signature of such reconnection event.
However, the outflow is less extended than the density structure, and solely present after the crossing of the HCS, whereas the density structures cover the whole HPS.
Thus, we may be far from the reconnection site so that a part of the outflow had time to slow down.

Furthermore, reconnection in the solar wind is usually identified assuming a Petschek-like scenario of reconnection
\citep[e.g.][]{Gosling_2005_SW_reconnection, Phan_2006_multi_sc_X-line, Eriksson_2022_HCS_avalanche_reconnection, Fargette_2023_reconnection_clustering_solo}, with rotational discontinuities along the separatrices of the exhaust, identified using the Walén relation \citep{Hudson_1970_anisotropic_RH}.
However, this event differs from reconnection between open field lines commonly reported in the near Sun HCS \citep{Phan_2020_PSP_reconnection_encounter1, Phan_2021_Reconnection_HCS, Phan_2024_HCS_FRs_merging}.
Indeed, tests of the Walén relation fail to unambiguously identify this event as a Petschek-like reconnection jet (see Appendix \ref{sec:appendix_Walen}).
More precisely, the Walén relation is only approximately satisfied, with a correlated Alfvén speed reduced by a factor 0.8 in the time interval $t \in [0, 0.4]$~h, while the observed velocity is far from  being alfvénic at later times.

A reconnection between open field lines of opposite polarities implies a local rearrangement of the magnetic field.
The reconnected field lines are connected to the Sun at both ends sunward of the reconnection site, and disconnected from the Sun anti-sunward of the reconnection site.
The observed outflow here is in the $+R$ direction, which, if it comes from a reconnection event, indicates that PSP is anti-sunward of the reconnection site.
However, there is no coincident dropout of strahl electrons (assumed to mostly be of solar origin) as expected on field lines disconnected from the Sun \citep[e.g.][]{Gosling_2005_HCS_reconnection_dropout, Phan_2021_Reconnection_HCS}.
It is therefore unlikely that the plasma outflow and density structure are due to a nearby magnetic reconnection event between open field lines.

Another possible origin of the density structure would be pinch-off magnetic reconnection occurring above the tip of an helmet streamer \citep{Wang_2000_streamer_blobs, Réville_2020_Tearing_Simu_PDS_HCS-HPS, Poirier_2023_HCS_WISPR_simu-obs}.
As presented in \citet{Réville_2022_flux-rope_HCS} with 3D MHD simulations, a thermal instability followed by a 3D tearing mode reconnection could lead to the creation of flux ropes, with scales comparable to those of the HPS, retaining a connection to the Sun.
However there is no signature of flux rope, with a scale comparable to the outflow, in the near Sun magnetic field measured by PSP (Figure \ref{fig:walen}(a), Appendix \ref{sec:appendix_Walen}).
This scenario is therefore also not able to explain the generation of the observed structure.

\subsection{Generation by interchange reconnection}

A possible explanation is that the observed plasma structuring is due to interchange reconnection \citep{Crooker_2002_interchange_definition_CMEs}, driven by photospheric motions as in \citet{Higginson_2017_interchange_CHB_dynamics, Aslanyan_2022_streamer_pseudostreamer_dynamic_coupling}, and released in the solar wind in a process similar to what has already been postulated by \citet{Wang_1998_blobs_HCS-HPS, Wang_2000_streamer_blobs, Crooker_2004_HPSs}.
Let's consider the following scenario, for which the initial configuration is schematized on the left of Figure \ref{fig:schema_interchange}.
An initial current sheet (transparent blue shape in Figure \ref{fig:schema_interchange}) lies between the open and closed magnetic field lines below the streamer stalk.
Random photospheric motion of the solar plasma are dragging the magnetic lines footpoints.
This induces changes in the whole streamer's magnetic structure.
The initial current sheet is progressively getting thinner, and eventually rendered unstable, triggering magnetic reconnection between the open and closed field lines.
The hotter and denser plasma initially confined within the coronal closed loops is then released in the new open field, building a dense solar wind outflow \citep{Krasnoselskikh_2023_interchange_reconnection}.

The measured density substructures would be linked to inhomogeneities already existing in the solar corona, which constitutes a network of loops with different plasma characteristics.
The injected plasma is therefore more or less dense depending on the reconnected loop.
As interchange reconnection continues to operate, the plasma from each loop is then injected one after the other in the solar wind open field lines.

A first reconnection event between a coronal loop and an open field line of a certain polarity would induce a rapid change in the newly formed loop.
Moreover, due to the continuous photospheric motion, the current sheet also extends to the other side of the helmet streamer.
The change of topology, due to the magnetic reconnection event, would communicate to this other side in a time scale $\tau_A = L/V_A \sim 1$~min, inferred from a characteristic length $L \sim 10^5$~km and $V_A \sim 2 \times 10^3$~km/s (a characteristic coronal Alfvén speed).
As the observed density structure overlaps the two magnetic sectors, the reconnection is expected to happen near the top of the streamer.
Thus, $L$ is taken as an order of magnitude lower than the total loop length scale ($\sim 10^6$~km), as $L$ only represents the length (along the loop) separating the two sides of the current sheet.
The first reconnection event rapidly disturbs and makes unstable the other side of the current sheet, triggering reconnection again, and so on.
This initiates a back and forth dynamic between the open field lines of opposed polarities.

In the described scenario, the interchange reconnection is expected to occur at the tip of the helmet streamer (red stars in Figure \ref{fig:schema_interchange}), so with the most prominent loop.
The time spent extracting plasma from one coronal loop ($\tau_{\rm rec}$), would have the same timescale as the observed density substructuring (circled green numbers in Figure \ref{fig:structure_densite} (a)), so $\tau_{\rm rec} \sim 20$~min.
This leads to $\tau_{\rm rec} \gg \tau_A$, such that the back and forth reconnection between the loop and open field lines of the two polarities is a quasi-steady process.
In consequence, both sides of the streamer are expected to interact with the same coronal loop.
Therefore, the plasma from this loop would simultaneously be injected in the open field lines of both magnetic polarities, which explains the coherence of the density structure over the HPS.

The longitudinal extension of the magnetic field inversion is $\sim 2$° (Section \ref{sec:density_structure_HPS-HCS}), which translates to a source region of $\sim 2.5 \times 10^4$~km at 2 solar radii.
This order of magnitude is however probably overestimated as the reconnection outflows are always larger than the reconnecting region.
The reconnection region is therefore most probably smaller than what expected from the angular extensions of the HPS in the heliosphere.



\section{\label{sec:conclusion}Discussion and conclusion}

To summarize, a radial alignment between PSP and SolO around the 29/04/2021 allowed us to find the same density structure (Figure \ref{fig:structure_densite}), which is part of the HPS (Figure \ref{fig:HCS-HPS_PSP-SolO}), passing through the two spacecraft after a 137.6~h propagation time from 0.075 to 0.9~au.
The two spacecraft measurements of the density structure were better matching when compared as functions of time.
This implies that the identified density structure has mostly radial gradients.
Accounting for the increase of radial speed reveals a compression by a factor $\sim$ 1.3-1.5.
This compression can be explained by an SIR development from PSP to SolO (see Appendix \ref{sec:SIR}).

\begin{table}
\caption{\label{tab:structure_size}Estimations of the minimum density structure's sizes at 3 different distances from the Sun.}
\centering
\begin{tabular}{clll}
\hline \hline
& \multicolumn3c{Sizes (km)} \\
Distance to Sun & L$_R$ & L$_\varphi$ & L$_\theta$  \\
\hline
2-3 R$_S$ (source) & $\lesssim 10^{6}$ & 10$^{5}$ & $3 \times 10^{4}$ \\
0.075 au (PSP) & 10$^6$ & $9 \times 10^{5}$ & $2 \times 10^{5}$ \\
0.9 au (SolO) & $1.5 \times 10^6$ & $1.1 \times 10^{7}$ & $2 \times 10^{6}$ \\
\hline
\end{tabular}
\end{table}

The magnetic field measurements, at first, were seemingly indicating that the sector boundaries were crossed in opposite directions (Section \ref{sec:mag_inversion}), as the $B_R$ and the electron PA reversals were opposite between PSP and SolO (Figure \ref{fig:HCS-HPS_PSP-SolO} (b, f, g)).
Comparing measurements as functions of an angular extension $\Delta \phi$ in the Carrington reference frame (Figure \ref{fig:HCS-HPS_PSP-SolO} (a'-g')) shows reversals in the same direction.
This is because PSP and SolO orbits around the Sun are respectively faster and slower than the Sun rotation.
These large scale reversals are linked with the solar wind's source on the Sun.
Then, the $B_R$ sector pattern is the same when the data are set in the Carrington reference frame.

Moreover, the $B_R$ large scale inversions also extend over comparable longitude intervals ($\sim$2-3°, Figure \ref{fig:HCS-HPS_PSP-SolO}), which indicates that the sector boundaries have kept a similar angular width during the solar wind's expansion.
This is in agreement with a nearly spherical expansion.
The $B_R$ reversal region being slightly shorter at SolO could moreover be explained by the SIR formation and compression.

The displacement of the HCS and TSB as compared to the density structure from PSP to SolO, as well as the seemingly change of magnetic polarity measured by both spacecraft within the HPS, can be linked to the 3D configuration of the HPS.
The two spacecraft are located at different latitudes and longitudes, and also have different trajectory within the plasma.
A local tilt of the HCS, as compared to the density structures, could therefore be at the origin of this observed discrepancy.

In-situ observations are not concordant with a generation of the density structures by reconnection between solar wind open field lines, as there is no measured suprathermal electron flux dropout within the $+V_R$ outflow.
A generation by pinch-off reconnection at the tip of a helmet streamer due to tearing instability is also not in agreement with the present observations, as no signature of flux rope at the HPS's scale is observed at PSP.
The observed density substructures are instead interpreted as coronal loop plasma injected in the solar wind along open field lines through interchange reconnection \citep{Pontin-Priest_2022_review_MHD_reconnection_3D}.
This measured in-situ substructuring could then be due to the intrinsic structuring of the corona in dense loops, where consecutive interchange reconnection events predominantly operate with one loop at a time.
This scenario is in agreement with our observations of the low alpha abundance on both satellites. Indeed, low helium abundances can be a signature of a solar wind released by reconnection near the cusp of helmet streamers, where plasma is gravitationally stratified \citet{Yogesh_2024_helium_abundances}.

This density structure was found to have, in an inertial frame (HCI), a minimum longitudinal extension $\Delta \varphi_{\rm str} \sim 6$°, or $\sim 4$° when taking into account the non-radial propagation effects.
The minimal latitudinal extension of the structure $\Delta \theta_{\rm str}$ was inferred to be $\sim 3$° or $\sim 1$°, when considering radial and non-radial propagation, respectively \citep{Berriot_2024_propag}.

Based on the results discussed above, we now assume that the density structure is of solar origin (around 2-3 $R_s$), and that it keeps a constant angular extension during the nearly spherical expansion of the solar wind.
This allows to estimate the radial evolution of the overall structure's sizes, which is summarized in Table \ref{tab:structure_size}.
The structure minimum non-radial sizes in Table \ref{tab:structure_size} have then been estimated as L$_\varphi = R \, \Delta \varphi_{\rm str}$ and L$_\theta = R \, \Delta \theta_{\rm str}$, (with $R$ in km and $\Delta \varphi_{\rm str}, \Delta \theta_{\rm str}$ in radians), while the structure radial size near the corona is expected to be slightly smaller than at PSP due to the plasma's acceleration.
The angles used for the calculations are those estimated by taking into account non-radial deflections.
As can be seen from Table \ref{tab:structure_size}, what probably was originally an elongated structure in the radial direction near the corona  became vaguely spherical at PSP's distance, and evolved to be quite flat by SolO's orbit.
This expansion is depicted in Figure \ref{fig:schema_interchange} (b).
We do not expect large enough departures from the hypotheses used for the sizes estimations that they would change the qualitative structure's aspect ratio evolution.
A recent statistical analysis of periodic density structures near 1~au by \citet{Di-Matteo_2024_PDSs_azimuthal_sizes} also reported sizes more elongated in the transverse direction ($\sim 2 \times 10^6$~km) than in the radial one ($\sim 5 \times 10^5$~km).

The evolution of the identified plasma parcel seems similar to what is described by \citet{Plotnikov_2016_Tracking_of_Corotating_Density_Structures}, using both remote-sensing and in-situ data from the STEREO-A and STEREO-B spacecraft. They reported density inhomogeneities gradually being caught up by SIRs while advected with the surrounding slow solar wind.
These inhomogeneities were also found to be mainly originating from near the polarity inversion in the corona.

Our interpretation of the studied case is the following.
A density structure \citep[akin to the ones observed by][]{Sheeley_1997_blobs, Wang_1998_blobs_HCS-HPS, Wang_2000_streamer_blobs, Viall_2015_PDS, Sanchez-Diaz_2017_remote_blobs_HCS} emerges near the tip of a streamer through interchange magnetic reconnection.
Due to the dynamical processes at its origin, this structure is formed of 4 principal substructures and mostly has radial gradients.
This structure, which is a part of the HPS, then propagates in the inner heliosphere with the surrounding slow solar wind, not being destroyed and keeping its identity, from at least $\sim$0.075 to $\sim$0.9~au.
As it gets further away from the Sun, a SIR develops and takes over the slow wind with the HPS, also sweeping up the density structure (Appendix \ref{sec:SIR}).

A next study will focus on the finer scale HCS/HPS physics in order to more precisely characterize the plasma evolution between the two spacecraft.
The interface between the unperturbed slow wind and the SIR also exhibits complex plasma features, most likely due to the interaction between the HCS and SIR forward's edge.
This however requires a further analysis, which goes beyond the scope of the present study.


\begin{acknowledgments}
The authors acknowledge funding support from CNES.
We acknowledge the SWEAP team led by J. Kasper for use of SWEAP data, and the FIELDS team for use of FIELDS data (\hyperlink{https://doi.org/10.48322/0yy0-ba92}{https://doi.org/10.48322/0yy0-ba92}).
The FIELDS experiment on the Parker Solar Probe spacecraft was designed and developed under NASA contract NNN06AA01C.
Solar Orbiter is a mission of international cooperation between ESA and NASA, operated by ESA.
Solar Orbiter SWA data were derived from scientific sensors that were designed and created and are operated under funding provided by numerous contracts from UKSA, STFC, the Italian Space Agency, CNES, the French National Centre for Scientific Research, the Czech contribution to the ESA PRODEX program, and NASA.
Solar Orbiter SWA work at the UCL/Mullard Space Science Laboratory is currently funded by STFC Grant ST/X/002152/1.
We acknowledge the teams of SWA (\hyperlink{https://doi.org/10.5270/esa-ahypgn6}{https://doi.org/10.5270/esa-ahypgn6}), MAG (\hyperlink{https://doi.org/10.5270/esa-ux7y320}{https://doi.org/10.5270/esa-ux7y320}), and RPW (\hyperlink{https://doi.org/10.57780/esa-3xcjd4w}{https://doi.org/10.57780/esa-3xcjd4w}) for providing data used in this study.
\end{acknowledgments}

\appendix

\section{\label{sec:appendix_phi}Estimation of the density structure longitudinal extension}

\subsection{\label{sec:appendix_phi_r}Case of a radial propagation}

In order to estimate the minimal longitudinal extension of the structure $\Delta \varphi_{\rm str}$ for the studied case, we first consider a purely radial propagation of the plasma.
Therefore, in the studied case, $\Delta \varphi_{\rm str}$ comes from the addition of three effects:
\begin{equation}
     \Delta \varphi_{\rm str} = \Delta \varphi_{\rm SolO} + \Delta \varphi_{\rm PSP} + \Delta \varphi_{\rm shift}
\end{equation}
where $\Delta \varphi_{\rm PSP}$ and $\Delta \varphi_{\rm SolO}$ correspond PSP and SolO longitudinal scan of the density structure, respectively.
The term $\Delta \varphi_{\rm shift}$ is the longitude shift between the two spacecraft scans, as sketched in Figure \ref{fig:schematic_structures_inertial}.

The two contributions, $\Delta \varphi_{\rm sc}$, coming from the longitude scanned by the spacecraft during the crossing of the structure are:
\begin{equation}
    \Delta \varphi_{\rm SC} = \varphi_{\rm SC}(t=t_{\rm end, SC}) - \varphi_{SC}(t=t_{\rm start, SC}),
\end{equation}
where $\varphi_{\rm SC}(t)$ is the spacecraft SC (PSP or SolO) longitude.
The times at which the spacecraft enters and exits the structure are $t_{\rm start, SC}$ and $t_{\rm end, SC}$, respectively.
The longitude scanned by SolO ($\Delta \varphi_{\rm SolO} \sim 0.06$°) during the structure passage is much lower than the one scanned by PSP ($\Delta \varphi_{\rm PSP} \sim 3.4$°) as here $\omega_{\rm PSP} / \omega_{\rm SolO} \simeq 64$ where $\omega_{\rm PSP}$ and $\omega_{\rm SolO}$ are the angular rotation frequencies of PSP and SolO, respectively.

The last contribution, $\Delta \varphi_{\rm shift}$, comes from the longitude shift between the structure exit at PSP, and arrival at SolO, such that:
\begin{equation}
    \Delta \varphi_{\rm shift} = \varphi_{SolO}(t=t_{\rm start, SolO}) - \varphi_{PSP}(t=t_{\rm end, PSP}).
\end{equation}
This finally allows to express the minimum extension of the structure as:
\begin{equation}
    \Delta \varphi_{\rm str} = \varphi_{\rm SolO}(t=t_{\rm end, SolO}) - \varphi_{\rm PSP}(t=t_{\rm start, PSP}).
\end{equation}

\subsection{\label{sec:appendix_phi_nr}Correction for a non-radial propagation}

In the case of a non-radial propagation, the longitudinal deflection induces a $\Delta \varphi_{\rm NR}$ correction, such that, the minimal extension of the structure $\Delta \varphi_{\rm str}^*$ is:
\begin{equation}
    \Delta \varphi_{\rm str}^* = \Delta \varphi_{\rm str} - \Delta \varphi_{\rm NR},
\end{equation}
with
\begin{equation}\label{eq:phi_nr}
    \Delta \varphi_{\rm NR} = \varphi_p (t=t_{\rm out}) - \varphi_p (t=t_{\rm in}),
\end{equation}
where $\varphi_p (t)$ is the plasma longitude.
This deflection is:
\begin{equation}\label{eq:phi_nr_int_t}
    \Delta \varphi_{\rm NR} = \int_{t_{\rm in}}^{t_{\rm out}} \omega (t) dt = \int_{t_{\rm in}}^{t_{\rm out}} \frac{V_{\varphi}(t)}{R(t)} dt,
\end{equation}
where $t_{\rm in}$ and $t_{\rm out}$ are the structure passage time at the inner and outer spacecraft respectively. We suppose here that the variation of $\Delta \varphi_{\rm NR}$ during the structure passage at the two spacecraft can be neglected as compared to its variation during the propagation.
Since $V_R = \frac{dR}{dt} \Rightarrow dt = \frac{dR}{V_R}$, the non-radial correction can also be rewritten as:
\begin{equation}\label{eq:phi_nr_int_R}
    \Delta \varphi_{\rm NR} = \int_{R_{\rm in}}^{R_{\rm out}} \frac{V_{\varphi}(R)}{V_R(R)} \frac{dR}{R},
\end{equation}
where $R_{\rm in} = R(t=t_{\rm in})$ and $R_{\rm out} = R(t=t_{\rm out})$.
In the case of a constant propagation velocity ($V_R \; \& \; V_\varphi = \text{constant}$), this leads to
\begin{equation}\label{eq:phi_nr_cst_speed}
    \Delta \varphi_{\rm NR} = \frac{V_{\varphi}}{V_R} \int_{R_{\rm in}}^{R_{\rm out}} \frac{dR}{R} = \frac{V_{\varphi}}{V_R} \ln \left( \frac{R_{\rm out}}{R_{\rm in}} \right).
\end{equation}
In general, such as in the studied case, neither $V_R$ or $V_\varphi$ are constant.
Here, instead of solving analytically Equation (\ref{eq:phi_nr_int_R}), we will rather directly evaluate Equation (\ref{eq:phi_nr}) using results from the previous study.
The plasma longitude at $t_{\rm in}$ is by definition $\varphi_p (t=t_{\rm in}) = \varphi_{PSP} (t=t_{\rm in})$, and $\varphi_p (t=t_{\rm out})$ can be estimated using the propagation method with non-radial velocity \citep[][Appendix A]{Berriot_2024_propag}, such that here $\Delta \varphi _{\rm NR} \simeq 2.0$°.
Estimation of $\Delta \varphi_{\rm NR}$ using Equation (\ref{eq:phi_nr_cst_speed}), with $V_\varphi = 10$~km/s ($\sim V_T$) and $V_R = 250$~km/s averaged between PSP and SolO measurements, gives $\Delta \varphi_{\rm NR} = 5.7$°.
This is higher than the previous estimation for two reasons.
The first reason is that in the propagation method, $V_\varphi = V_T$, with a fixed RTN reference frame based on PSP's position when it crossed the plasma.
This therefore assumes a rectilinear non-radial deflection from PSP's position, with an angle depending on the ratio $V_\varphi/V_R$, while the analytical case assumes a rotation of the plasma.
The second reason is that the analytic expression includes a constant $V_\varphi$ at all distances, leading to a more important $\omega$ near the Sun.
However, since here the tangential deflection is due to the SIR, $V_\varphi$ appears further away, such that the estimation, obtained with the non-radial propagation of \citet{Berriot_2024_propag}, should be the most relevant.


\begin{figure*}[t!] 
\includegraphics[width=.9\textwidth]{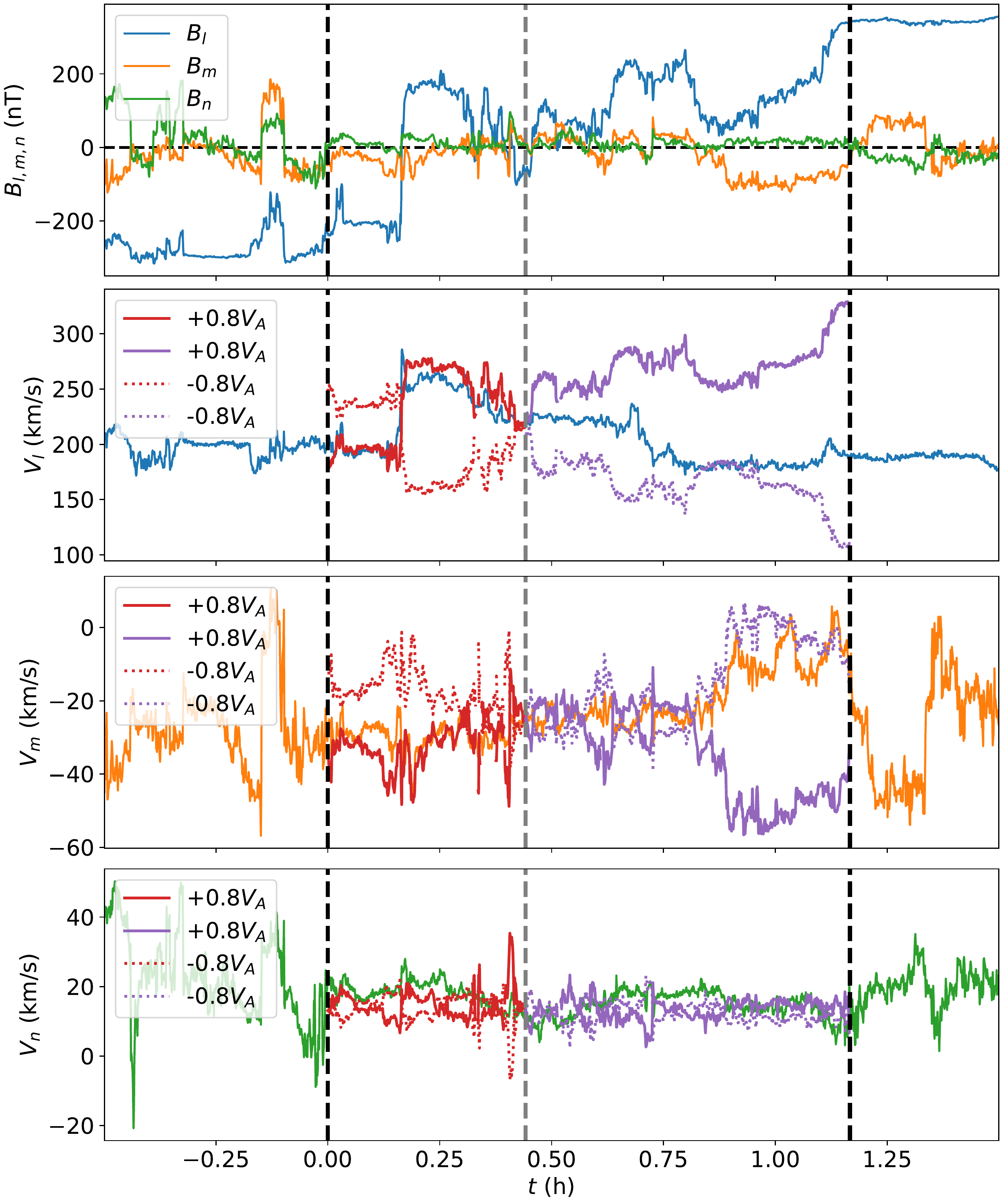}
\caption{\label{fig:walen}Test of the Walén relation along PSP crossing of the HPS.
Panel (a) shows the three components of the magnetic field in the $lmn$ reference frame of the minimum variance analysis, while panels (b), (c) and (d) show the $V_l$, $V_m$, and $V_n$ components of the proton bulk velocity, respectively.
Tests of the Walén relation for each velocity component are shown in panels (b,c,d), with red and purple lines.
The ($l,m,n$) reference frame is defined by computed the MVA between the two vertical black dashed lines.
The vertical grey dashed line marks the separation between the two sides of the outflow, see text for more details.
}
\end{figure*}  

\section{\label{sec:appendix_Walen}Test of the Walén relation on PSP data}


Magnetic reconnection in the solar wind is often identified using the Walén relation to check the alfvénicity of the reconnection outflow, which is expected to be bounded by two rotational discontinuities \citep{Gosling_2005_SW_reconnection, Phan_2020_PSP_reconnection_encounter1, Phan_2021_Reconnection_HCS, Eriksson_2022_HCS_avalanche_reconnection, Fargette_2023_reconnection_clustering_solo}.
This identification consists in comparing variations of the measured solar wind velocity $\Delta \mathbf{V}$ with
\begin{equation}\label{eq:walen}
 \pm \Delta \mathbf{V}_A = \pm \Delta \frac{\mathbf{B}}{\sqrt{\rho \mu_0}} \sqrt{1 - \frac{P_\parallel - P_\perp}{B^2/ \mu_0}},
\end{equation}
where $\mathbf{B}$ is the magnetic field vector, $\rho$ the plasma mass density, and $P_\parallel$ and $P_\perp$ the pressure parallel and perpendicular to $\mathbf{B}$, respectively.
The Alfvén velocity variations are defined as $\Delta \mathbf{V}_A =\mathbf{V}_A - \mathbf{V}_{A,0}$, with $\mathbf{V}_A$ and $\mathbf{V}_{A,0}$ the local and reference Alfvén velocity, respectively.
An outflow is identified as a reconnection exhaust when variations verify $\Delta \mathbf{V} = + \Delta \mathbf{V}_A$ on one side of the outflow, and $\Delta \mathbf{V} = - \Delta \mathbf{V}_A$ on the other, so that there is a jet with a velocity $\Delta \mathbf{V}_A$.
Considering that the solar wind velocity can be expressed as alfvénic fluctuations on top of a background plasma velocity $\mathbf{V} = \mathbf{V}_{bg} + \mathbf{V}_A$, velocity variations would, in general, include contributions from both $\mathbf{V}_A$ and $\mathbf{V}_{bg}$, so that  $\Delta \mathbf{V} = \Delta \mathbf{V}_{bg} + \Delta \mathbf{V}_A = (\mathbf{V}_{bg} - \mathbf{V}_{bg,0}) + (\mathbf{V}_A - \mathbf{V}_{A,0})$.
The identification using the Walén relation therefore assumes a constant background velocity $\Delta \mathbf{V}_{bg} = 0$. 
Moreover, this test is of particular importance at the boundaries of the jet, in order to check for the presence of rotational discontinuities \citep{Hudson_1970_anisotropic_RH}.

Observed reconnection outflows are usually slower than the estimation given in Equation (\ref{eq:walen}) due to potentially complex exhausts boundaries \citep[see for example][and references therein]{Phan_2021_Reconnection_HCS}.
To account for such effect, the Walén relation can then be considered to be verified if the velocity variations are such that $ \Delta \mathbf{V} = \pm \varepsilon \Delta  \mathbf{V_A}$, with $0 < \varepsilon \leq 1$, with $\varepsilon$ usually not far from $1$.

We show in Figure \ref{fig:walen} a test of the Walén relation for PSP's crossing of the HPS.
The magnetic field and velocity vectors have been rotated in the ($l,m,n$) reference frame, taken as the minimum variance reference frame of $\mathbf{B}$ \citep{Sonnerup_1967_MVA_historic, Sonnerup_1998_MVA} computed between the two vertical black lines.
For a planar current sheet, $l$ (maximum variance direction) corresponds to the direction of magnetic field reversal, $n$ (minimum variance direction) is the normal to the current sheet, and $m$ (intermediate variance direction) is the out of plane direction (along the guide field).
Here, the $l$ direction is almost parallel to the radial direction due to the dominance of the $B_R$ component, while the $m$ and $n$ directions are mostly lying in the $T$-$N$ plane.
This HPS structure is obviously more complex than a single theoretical planar current sheet, and is composed of several substructure of different scales, but the MVA still offers a privileged reference frame by separating well the l direction ($\lambda_l / \lambda_m = 7$ and $\lambda_l / \lambda_n = 66$).

On top of velocity measurements are shown predictions for the Walén relation, plotted as $\mathbf{V}_0 \pm \varepsilon \Delta \mathbf{V}_A = \mathbf{V}_0 \pm \varepsilon (\mathbf{V}_A - \mathbf{V}_{A,0})$.
The reference velocities $\mathbf{V}_{0}$ and $\mathbf{V}_{A,0}$ are taken as the velocities values at times corresponding to the grey vertical dashed line, separating the two sides of the outflow.
Plain and dotted curves are drawn for $\mathbf{V}_0 + \varepsilon \Delta \mathbf{V}_A$ and $\mathbf{V}_0 - \varepsilon \Delta \mathbf{V}_A$ respectively, with $\varepsilon = 0.8$.
The anisotropy factor $P_\parallel$ - $P_\perp$ has been estimated from the proton pressure only.

The left side of the outflow has a good correspondence between the bulk plasma velocity and the prediction (plain red curves).
The vertical grey dashed line, separating the two sides of the outflow (extending from $t \sim 0.15$~h to $t \sim 0.75$~h), has been chosen to be approximately where the velocity variations start to depart from correlated alfvénic fluctuations.
On the right side of the outflow, the large scale fluctuations in the $m$ direction seem to be reasonably well anti-correlated with the Alfvén speed for the chosen $\varepsilon = 0.8$, while fluctuations in the $n$ direction are inconclusive given their small amplitude.
However, for the $l$ direction (which should be the outflow direction for a reconnecting current sheet), neither $\mathbf{V}_0 \pm \varepsilon \Delta \mathbf{V}_A$ curves are in agreement with the measured plasma velocity.
The large discrepancy between the expected and measured plasma velocity lead us to conclude that this outflow most probably does not originate from the reconnection of a nearby solar wind X-line.
In addition, there is no evidence for the presence of a rotational discontinuity bounding the right side of the outflow.
If this part of the outflow comes from a reconnection jet, it was much decelerated at the time is crosses PSP.

It should also be noted that the temporal scales of solar wind reconnection exhausts identified with this method are usually smaller than the one of the observed plasma jet \citep[e.g.][]{Gosling_2006_Helios_reconnection, Phan_2021_Reconnection_HCS, Eriksson_2022_HCS_avalanche_reconnection, Eriksson_2024_PSP_reconnection}.
Smaller scale reconnection event might exist in this HPS but, due to their sizes, they would not directly be related to the density substructures generation.
In fact, a small scale reconnection event (of duration $\lesssim 1$~min) near this particular HCS has already been identified in a previous study, around $t = 10$~min ($= 29/04/2021$~00:55~UTC, see the list given in \citet{Eriksson_2024_PSP_reconnection}).
Characterization of the HPS's finer scale dynamics however goes beyond the scope of the present study.


\begin{figure*}[t!] 
\resizebox{\hsize}{!}{\includegraphics{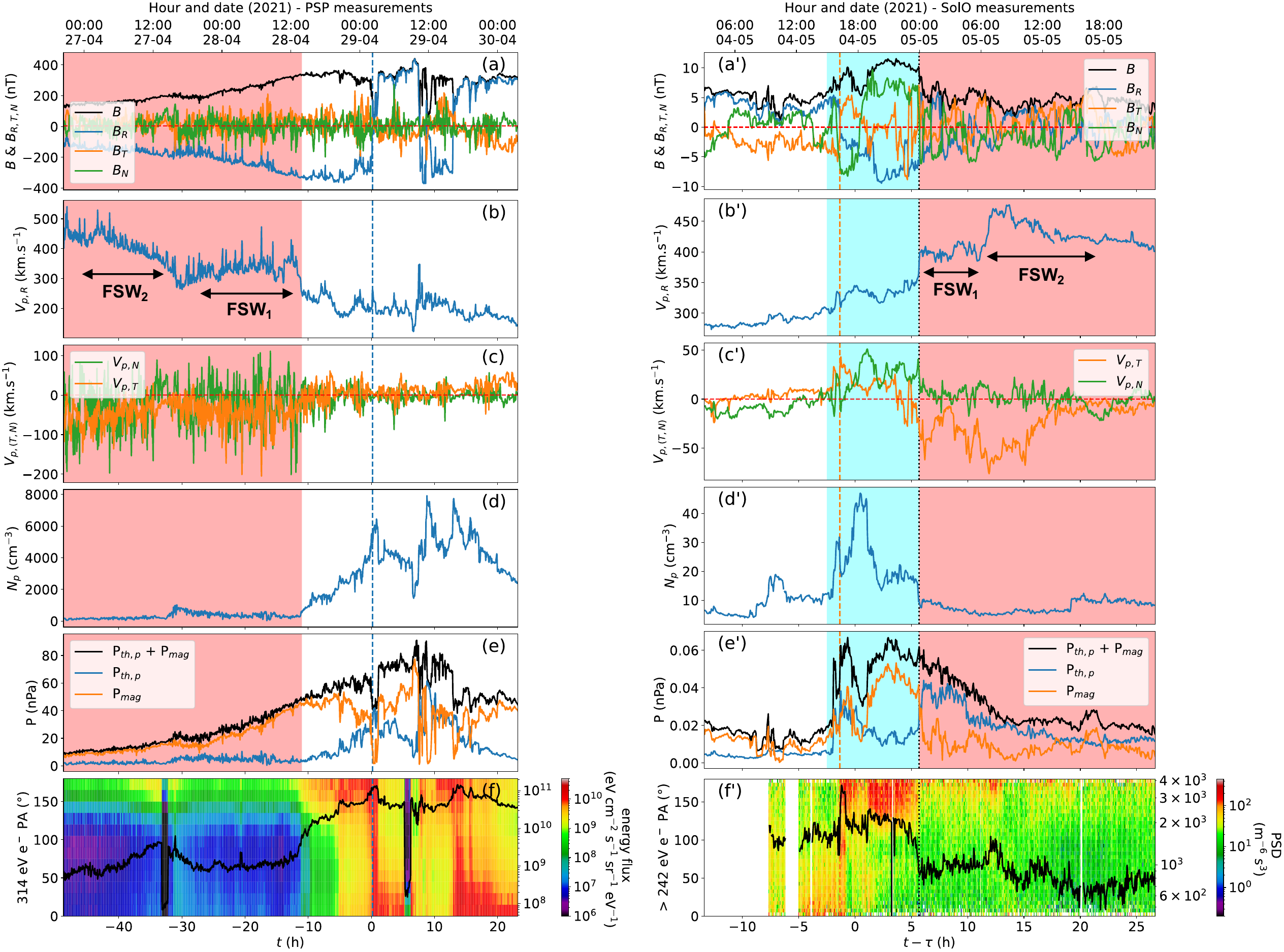}}
\caption{\label{fig:SIR}
Large scale overview of measurements  of PSP (left) and SolO (right) near HCS crossings (blue and orange vertical dashed lines).
Faster solar wind streams are highlighted in red for both spacecraft. The double arrows and "FSW$_{1,2}$", in panels (b) and (g), denote the two different sources of the faster wind.
The compressed slow wind on SolO is highlighted in blue, and the SIR's stream interface is marked by the black dotted vertical line.
Panels (a,a') show the magnetic field's magnitude (black) and its components (colors) in the RTN local frame,
(b,b') V$_{p,R}$ the radial component of the proton velocity vector, (c,c') the T and N components of the proton velocity, (d,d') the proton density, (e,e') the proton plus magnetic (black), proton (blue) and magnetic (orange) pressures, (f,f') the strahl electron PA energy flux.
}
\end{figure*}  

\section{\label{sec:SIR}Formation of a stream interaction region from PSP to SolO}

Stream interaction regions (SIRs) have been frequently observed in the solar wind \citep[see][and references therein]{Gosling_Pizzo_1999_CIRs_formation,Richardson_2018_SIR_review}.
In the studied case, an SIR has developed during the plasma propagation, as the slow wind encompassing the HCS and HPS has been caught up by a faster wind source.
Precedent studies already highlighted the close association between SIRs and the HCS \citep{Borrini_1981_HCS_Helium_SIR, Thomas_1981_HCS_eaten_CIR, Gosling_Pizzo_1999_CIRs_formation, Huang_2016_HCS_SIR, Jian_2019_SIR-HCS_relationship}.
Moreover, density structures such as those observed here have also been reported to sometimes be swept up by SIRs
\citep{Rouillard_2010_transients_remote, Rouillard_2010_transients_in-situ, Plotnikov_2016_Tracking_of_Corotating_Density_Structures}.

We show on Figure \ref{fig:SIR} a set of physical parameters measured by PSP (left panels) and by SolO (right panels) around the studied time intervals.
We indicated by vertical dashed lines (blue for PSP, orange for SolO) the position of the HCS studied previously (Figure \ref{fig:HCS-HPS_PSP-SolO}).
We recall (Section \ref{sec:Carrington_longitude}) that during the studied time intervals PSP is, as opposed to SolO, in super-rotation as compared to the Sun's surface.
Therefore, the two spacecraft are crossing solar wind streams in opposite directions.
From Figure \ref{fig:SIR}, one can see that an SIR has formed during the solar wind propagation from PSP to SolO.
A faster solar wind source (highlighted in red) has caught up and compressed the slower wind's region ahead (highlighted in blue in panels (a'-e')).
The density structure and HCS shown in Figures \ref{fig:structure_densite} and \ref{fig:HCS-HPS_PSP-SolO} are parts of this compressed slow wind, and were therefore also engulfed by the SIR.

Figure \ref{fig:SIR} (a-f) shows PSP observations between 27/04/2021~00:00~UTC and 31/04/2021~00:00~UTC ($\sim t \in [-49, 23]$~h).
The magnetic field intensity is relatively smooth, except for some HCS and HPS crossings. 
After the HCS studied above (near $t=0$~h), the $B_R$ component has indeed two more reversals along with simultaneous increase in density, and higher $\beta$ ($P_{th, p} > P_{mag}$).
There is also an overall change (mostly increase) of the magnetic field intensity, as expected due to the changing (mostly shortening) distance between PSP and the Sun.
Moreover, PSP measured two different faster solar wind streams ("FSW$_1$" and "FSW$_2$" in panel (b)), with speeds between $\sim$350-400~km/s around $t = -40$~h and between $\sim$300-350~km/s around $t = -20$~h.
These faster winds are presently far enough from the slow wind stream in the region of the studied HCS (dashed line) and HPS, so that these structures are still unperturbed.

On Figure \ref{fig:SIR} panels (a'-f') are SolO measurements between 04/05/2021~05:00~UTC and 05/05/2021~21:00~UTC ($\sim t - \tau \in [-13, 27]$~h), where the HCS is marked by the vertical orange dashed line.
The region highlighted in blue ($\sim t - \tau \in [-2, 6]$~h), presents an increase of magnetic field intensity $B$, proton density, proton pressure, and strahl electrons energy flux.
Within this region, the measured plasma also has mostly positive non-radial velocity components $V_{p,T}$ and $V_{p,N}$, and a larger radial velocity $V_{p,R}$ than the slow wind ahead.
We hence interpret the blue region as a compression and deflection of the slow wind caught up by the faster solar wind (highlighted in red).
This compression also brings the magnetic field lines closer to each other, increasing both the measured energy flux of the strahl electrons traveling along them, and the magnetic field intensity.
This also explains why the HPS longitude extension is slightly shorter at SolO than at PSP, see Section \ref{sec:mag_inversion}.

Next, the magnetic field magnitude has a prominent local minima corresponding to the identified density structure (around $t - \tau = 0.5$~h, Figures \ref{fig:structure_densite} and \ref{fig:HCS-HPS_PSP-SolO}).
Behind, so for larger $t - \tau$ values, two faster streams are present, as already described at PSP.
Due to the interaction with the slow solar wind ahead, these faster streams are deflected in the -T direction and present an increasing pressure when closer to the slow wind.
We precise the stream interface separating the slow and faster winds by the black vertical dotted line in the right panels of Figure \ref{fig:SIR}.
This stream interaction is more complex than a fast-slow stream interaction since it involves also the interaction of two fast streams.
The wind source denoted FSW$_1$ has been caught up by an even faster wind stream FSW$_2$. Therefore, FSW$_2$ is  deflecting FSW$_1$ wind in the +T direction, in opposition with the deflection in the $-$T direction induced by the interaction with the slow wind ahead.
This is likely the origin of the complex behavior of $V_T$ observed within FSW$_1$ in SolO measurements.

In order to understand this large scale data, we propose the following scenario.
After some time, the faster solar wind progressively catches up the slow wind carrying the density structure and associated HPS and HCS.
At the stream interface, this creates a pair of forward and reverse waves propagating respectively in the slow and fast winds, compressing and deflecting them \citep[e.g.][]{Gosling_Pizzo_1999_CIRs_formation}.
The SIR eventually engulfs the HCS and HPS because of the forward wave propagation.
This wave also steepens with time, leading to the large scale pressure discontinuity observed at SolO at $t - \tau = -2$ h in panel (e').  This marks the front limit of the SIR.
We note that this interface seems quite complex and coincides with the border of a smaller scale structure flanking the HCS.

The most remarkable part is that despite the solar wind evolution and development of the interaction region, not only the density structure and associated substructures have not been destroyed, but they are also still identifiable when comparing the measurements of both spacecraft (see Figure \ref{fig:structure_densite}).
The formation of a SIR furthermore explains the observed non-radial density structure compression, leading to a density $\sim$ 1.3-1.5 times higher (see Section \ref{sec:plasma_compression}) than what expected from a purely spherical expansion with the observed solar wind acceleration (Section \ref{sec:context}, Figure \ref{fig:structure_densite}).

\bibliography{references}{}
\bibliographystyle{aasjournal}

\end{document}